\RequirePackage{ifpdf}
\ifpdf 
\documentclass[pdftex]{sigma}
\else
\documentclass{sigma}
\fi

\usepackage{mathrsfs}

\DeclareMathAlphabet{\mathpzc}{OT1}{pzc}{m}{it}

\def\on#1#2{\mathop{\vbox{\ialign{##\crcr\noalign{\kern2pt}
$\scriptstyle{#2}$\crcr\noalign{\kern2pt\nointerlineskip}
\kern-2pt$\hfil\displaystyle{#1}\hfil$\crcr}}}\limits}

\def\nn{ \nonumber }
\def\bq{ \begin{equation} }
\def\eq{ \end{equation} }
\def\ben{ \begin{eqnarray} }
\def\en{ \end{eqnarray} }

\numberwithin{equation}{section}

\begin{document}

\allowdisplaybreaks
\renewcommand{\PaperNumber}{097}

\FirstPageHeading

\renewcommand{\thefootnote}{$\star$}

\ShortArticleName{On the Darboux--Nijenhuis Variables for the Open
Toda Lattice}

\ArticleName{On the Darboux--Nijenhuis Variables\\
for the Open Toda Lattice\footnote{This paper is a contribution to
the Vadim Kuznetsov Memorial Issue ``Integrable Systems and
Related Topics''. The full collection is available at
\href{http://www.emis.de/journals/SIGMA/kuznetsov.html}{http://www.emis.de/journals/SIGMA/kuznetsov.html}}}

\Author{Yuriy A. GRIGORYEV and Andrey V. TSIGANOV}
\AuthorNameForHeading{Yu.A.  Grigoryev and A.V. Tsiganov}
\Address{St.Petersburg State University, St.Petersburg, Russia}
\Email{\href{mailto:com974@mail.ru}{com974@mail.ru},
\href{mailto:tsiganov@mph.phys.spbu.ru}{tsiganov@mph.phys.spbu.ru}}

\ArticleDates{Received November 17, 2006; Published online
December 30, 2006}

\Abstract{We discuss two known constructions proposed by Moser and
by Sklyanin of the Darboux--Nijenhuis coordinates for the open
Toda lattice.}


\Classification{37K10}

\section{Introduction}

A bi-Hamiltonian  manifold $M$ is a smooth manifold endowed with
two compatible bi-vec\-tors~$P$,~$P'$
 such that
\[
[P,P]=[P,P']=[P',P']=0,
\] where $[\cdot,\cdot]$ is the Schouten bracket.
The bi-vectors $P$, $P'$ determine a pair of compatible Poisson
brackets on $M$, for instance \bq \label{pois-br}
 \{f(z),g(z)\}=\langle df, P dg \rangle=\sum_{i,j}^{\dim M}
P^{ij}(z)\dfrac{\partial f(z)}{\partial z_i}\dfrac{\partial
g(z)}{\partial z_j}, \eq and similar brackets $\{\cdot,\cdot\}'$
to $P'$.

Dynamical systems on $M$ having enough functionally independent
integrals of motion $H_1,\ldots$, $H_{n}$ in involution with
respect to both Poisson brackets \bq\label{bi-ham}
\{H_i,H_j\}=\{H_i,H_j\}^\prime=0 . \eq will be called
bi-integrable systems.

A suf\/f\/icient condition in order that integrals of motion
$H_1,\ldots, H_{n}$ be in bi-involution is that the corresponding
vector f\/ields $X_{H_i}$ are bi-Hamiltonian vector f\/ields
\cite{ksm90,mag97}, which form a~so-called anchored Lenard--Magri
sequence \bq\label{len-mag} PdH_1=0,\qquad
X_{H_i}=PdH_i=P'dH_{i-1},\qquad P'dH_n=0. \eq

The class of manifolds we will consider are particular
bi-Hamiltonian manifolds, to be termed $\omega N$-manifolds, where
one of the two Poisson bi-vectors is nondegenerate  (say $P$) and
thus def\/ines a symplectic form $\omega=P^{-1}$ and, together with
the other one, a recursion operator \cite{fmp01,fp02}
\bq\label{recc-op} N=P^{ \prime}P^{-1}, \eq
 and its dual $ N^*=P^{-1}P'$. Operator $N$ is called
a Nijenhuis operator \cite{ksm90,mag97} or hereditary operator
\cite{dor79,ff81}.

One of the main property of the recursion operator $N$ is that
bi-vectors $P=P^{(0)}$ and $P'=P^{(1)}$ belong to a whole family
of compatible Poisson tensors \bq \label{mult-poi} P^{(k)}=N^k P ,
\eq which def\/ine a family of compatible Poisson brackets $\{\cdot
,\cdot\}^{(k)}$ on the $\omega N$-manifold $M$.

Another useful property  of $N$ is that normalized traces of the
powers of $N$ are integrals of motion satisfying Lenard--Magri
recurrent relations (\ref{len-mag}) \cite{mag97}:
\bq\label{mag-int} H_j=\frac{1}{2j}\, \mbox{\rm tr}\, N^j. \eq The
class of coordinates, called Darboux--Nijenhuis coordinates, are
canonical with respect to $\omega$ and diagonalize  recursion
operator $N$. According to \cite{fmp01,fp02}, the $n$-tuple
($H_0,\ldots,H_n$) of Hamiltonians on $M$ (where $n =\frac12 \dim
M$) is separable in Darboux--Nijenhuis coordinates if and only if
they are in involution with respect to both Poisson brackets
(\ref{bi-ham}).

In this paper we compare two known families of the separated
variables for the open Toda lattice \cite{mos75,skl85a} with the
corresponding Darboux--Nijenhuis coordinates.

\section{The separation of variables method}
\setcounter{equation}{0}

In the separation of variables method variables we are looking for
complete integral $S(q,t,\alpha_1,\ldots$, $\alpha_n)$ of the
Hamilton--Jacobi equation \bq\label{Eq-HJt} \dfrac{\partial
S}{\partial t}+H\left(q,\dfrac{\partial S}{\partial q},t\right)=0
,\qquad \det\left\|\dfrac{\partial ^2  S}{\partial q_i\partial
\alpha_j}\right\|\neq 0 , \eq
 where $q=(q_1,\ldots,q_n)$,  in the additive form
 \bq\label{Add-Int}
S(q,t,\alpha_1,\ldots,\alpha_n)= -Ht+\sum_{i=1}^n
S_i(q_i,\alpha_1,\ldots,\alpha_n). \eq Here the $i$-th component
$S_i$ depends only on the $i$-th coordinate $q_i$ and $n$
parameters $\alpha_1,\ldots,\alpha_n$ which are the values of
integrals of motion. In such a case $H$ is said to be separable
and coordinates $q$ are said to be separated coordinates for $H$,
in order to stress that the possibility to f\/ind an additive
complete integral of (\ref{Add-Int}) depends on the choice of the
coordinates.

For any complete integral of the equation (\ref{Eq-HJt}) solutions
$q_i=q_i(t,\alpha,\beta)$ and $p_i=p_i(t,\alpha,\beta)$ of the
Hamilton equations of motion are found from the Jacobi equations
\bq \label{Eq-J} \beta_i=-\dfrac{\partial  S}{\partial
\alpha_i},\qquad p_i=\dfrac{\partial S}{\partial q_i} ,\qquad
i=1,\ldots,n. \eq In the  separation of variables  method  the
each second Jacobi equation \bq\label{Eq-Js} p_i=\dfrac{\partial
}{\partial q_i}S_i(q_i,\alpha_1,\ldots,\alpha_n) , \eq contains
the pair of the Darboux coordinates $p_i$ and $q_i$ only. These
equations and their more symmetric form{\samepage \bq\label{Eq-S}
\Phi_i(q_i,p_i, \alpha_1,\ldots,\alpha_n)=0,\qquad
\det\left\|\dfrac{\partial \Phi_j}{\partial \alpha_k}\right\|\neq
0 , \eq are called the separated equations.}

The separated variables $(p,q)$ are def\/ined up to canonical
transformations $p_i\to f_i(p_i,q_i)$ and $q_i\to g_i(p_i,q_i)$
and integrals of motion $H_j=\alpha_j$ can be always  replaced
with $\tilde{H_j}=\phi_j(H_1,\ldots,H_n)$. Such transformations
change the form of the separated equations (\ref{Eq-S}).

For example, we can use canonical transformation to the
action-angle variables \bq\label{t2-aav} I_j=\frac{1}{2\pi}\oint
p_j dq_j=\frac{1}{2\pi}\oint \dfrac{\partial
S_j(q_j,\alpha_1,\ldots,\alpha_n)}{\partial q_j} dq_j ,\qquad
w_j=\frac{\partial W}{\partial I_j} . \eq Here
$W=\sum\limits_{i=1}^n S_i(q_i,\alpha_1,\ldots,\alpha_n)$ is
Hamilton's characteristic function. The action-angle
variab\-les~$I$,~$w$ are the special separated variables which
allow us to
 linearize equations of motion
\[
\dot{I}_j=0,\qquad \dot{w}_j=F_j(I_1,\ldots,I_n),\qquad
j=1,\ldots,n.
\]
However, the action-angle variables are not always convenient, for
instance for the quantum integrable system.

\begin{example}
Consider a two-particle open Toda chain with the following
integrals of motion \bq\label{tod2-int} H_1= p_1+p_2,\qquad
H_2=\frac{p_1^2+p_2^2}{2}+e^{q_1-q_2}. \eq Variables
\bq\label{2tod-xy} v_{1,2}=\frac{1}{\sqrt{2}}(q_1\pm q_2),\qquad
u_{1,2}=\frac{1}{\sqrt{2}}(p_1\pm p_2) \eq are separated
variables, because substituting these variables in the def\/inition
of integrals of motion we obtain desired separated equations
(\ref{Eq-S})
\[
\Phi_1(v_1,u_1,\alpha)=\sqrt{2} u_1-H_1=0,\qquad
\Phi_2(v_2,u_2,\alpha)
=e^{\sqrt{2}v_2}+\frac{u_2^2}{2}-H_2+\frac{H_1^2}{4}=0 .
\]
Provided  $H_i=\alpha_i$ and $u_i=\dfrac{\partial S_i}{\partial
v_i}$, these separated equations
\[
\frac{\partial S_1}{\partial v_1}=\frac{\alpha_1}{\sqrt{2}},\qquad
\frac12\left(\frac{\partial S_2}{\partial
v_2}\right)^2+e^{\sqrt{2}v_2}-\alpha_2+\frac{\alpha_1^2}4=0,
\]
can easily be integrated by quadratures
\begin{gather}
S_1=\frac{\alpha_1v_1}{\sqrt{2}},
\label{tod2-act}\\
S_2=\mp\sqrt{4\alpha_2-\alpha_1^2-4e^{\sqrt{2}v_2}}\pm
\sqrt{4\alpha_2-\alpha_1^2}
\mathrm{arctanh}\left(\frac{\sqrt{4\alpha_2-\alpha_1^2-4e^{\sqrt{2}v_2}}}
{\sqrt{4\alpha_2-\alpha_1^2}}\right) .\nn
\end{gather}
Then we have to substitute the corresponding Hamilton principal
function $S=-\alpha_2t+S_1+S_2$ in the Jacobi equations
\begin{gather*}
\beta_1=-\frac{\partial S}{\partial
\alpha_1}=-\frac{v_1}{\sqrt{2}}+\frac{\alpha_1}{\sqrt{4\alpha_2-\alpha_1^2}}
\,\mathrm{arctanh}\left(\frac{\sqrt{4\alpha_2-\alpha_1^2-4e^{\sqrt{2}v_2}}}
{\sqrt{4\alpha_2-\alpha_1^2}}\right),\\
\beta_2=-\frac{\partial S}{\partial \alpha_2}=t-
\frac{2}{\sqrt{4\alpha_2-\alpha_1^2}} \,\mathrm{arctanh}
\left(\frac{\sqrt{4\alpha_2-\alpha_1^2-4e^{\sqrt{2}v_2}}}
{\sqrt{4\alpha_2-\alpha_1^2}}\right) ,
\end{gather*}
and solve the obtained equations with respect to $v_{1,2}$
\begin{gather*}
v_1=\frac{\alpha_1(t-\beta_2)}{\sqrt{2}}+\sqrt{2}\beta_1,\\
v_2=\sqrt{2}
\ln\left(\alpha_2-\frac{\alpha_1^2}{4}\right)-2\sqrt{2}\ln\left(\cosh\left(\sqrt{\alpha_2-\frac{\alpha_1^2}4}(t-\beta_2)\right)\right)
.
\end{gather*}
Using these solutions and canonical transformation (\ref{2tod-xy})
we obtain closed equations for trajectories of motion in the
original variables \bq\label{tod2-feq}
q_{1,2}=\frac{\alpha_1(t-\beta_2)}{2}
\pm\frac{1}{2}\ln\left(\alpha_2-\frac{\alpha_1^2}4\right)
\mp\ln\left(\cosh\left(\sqrt{\alpha_2-\frac{\alpha_1^2}4}(t-\beta_2)\right)\right)-\beta_1.
\eq In similar way we can get $p_{1,2}(t)$ from the second Jacobi
equations (\ref{Eq-Js}).

Parameters $\beta_{1,2}$ may be excluded by the shifts $t\to
t+\beta_2$ and $q_i\to q_i+\beta_1$ and, thus, the solutions
$q_{1,2}(t)$ depend  on two values of integrals of motion
$\alpha_{1,2}$ only.

Now from (\ref{t2-aav}) one can easily f\/ind the action variables
\bq\label{tod2-act+} I_1=\alpha_1=\sqrt{2}u_1,\qquad
I_2=4\alpha_2-\alpha_1^2=2u_2^2+4e^{\sqrt{2}v_2} \eq and the angle
variables \bq\label{tod2-ug} w_1=\frac{\partial S}{\partial
I_1}=\frac{v_1}{\sqrt{2}},\qquad w_2=\frac{\partial S}{\partial
I_1}= -\frac{1}{2\sqrt{2}}\frac{
\mathrm{arctanh}\left(\frac{u_2}{\sqrt{u_2^2+2e^{\sqrt{2}v_2}}}\right)}
{\sqrt{u_2^2+2e^{\sqrt{2}v_2}}} . \eq In these coordinates
symplectic form has a standard form
\[\omega=\sum_{j=1}^2 dp_j\wedge dq_j=\sum_{j=1}^2 dI_j\wedge d w_j\]
and equations of motion are linearized
\[
\dot{I_j}=0,\qquad \dot{w_1}=\frac{\partial H}{\partial
I_1}=\frac{I_1}{2},\qquad \dot{w_2}=\frac{\partial H}{\partial
I_2}=\frac{1}{4} .
\]
Using solutions of these equations and inverted canonical
transformation $I,w\to p,q$ one can easily derive equations for
the trajectories of motion (\ref{tod2-feq}).

Separated variables $v_{1,2}$ and $u_{1,2}$ can also be used in
quantum mechanics, for example, to f\/ind the spectrum of
Hamiltonian of a periodical two-particle Toda chain
\cite{komts88}.

\end{example}

\subsection[Darboux-Nijenhuis coordinates]{Darboux--Nijenhuis coordinates}
In this section we will describe a class of canonical coordinates
on $\omega N$-manifolds, called Darboux--Nijenhuis coordinates.
They will play the important role of variables of separation for
(suitable) systems on $\omega N$-manifolds.

By def\/inition a set of local coordinates $(x_i, y_i )$ on an
$\omega N$-manifold is called a set of Darboux--Nijenhuis
coordinates if they are canonical with respect to the symplectic
form
\[\omega=P^{-1}=\sum_{i=1}^n dy_i\wedge dx_i
\]
and put the recursion operator $N$ in diagonal form,
\bq\label{xy-pedr}
 N=\sum_{i=1}^n\lambda_i\left(
\dfrac{\partial }{\partial  x_i}\otimes dx_i+ \dfrac{\partial
}{\partial y_i}\otimes dy_i\right). \eq This means that the only
nonzero Poisson brackets are \bq\label{br-xy}
\{x_i,y_j\}=\delta_{ij},\qquad \{x_i,y_j\}'=\lambda_i\delta_{ij}.
\eq The distinguishing property of the pairs of Darboux--Nijenhuis
coordinates $(x_i, y_i )$ is that their dif\/ferentials span an
eigenspace of $N^*$, that is, satisfy the equations
\bq\label{qv-pedr} N^*dx_i=\lambda_idx_i,\qquad
N^*dy_i=\lambda_idy_i \eq As a consequence of the compatibility
between  $P$ and $P'$, the Nijenhuis torsion of $N$ vanishes
\bq\label{kr-n} T_N(X,Y)=[NX,NY]-N\bigl( [NX,Y]+[X,NY]-N[X,Y]
\bigr)=0, \eq here  $X$, $Y$ are  arbitrary vector f\/ields on $M$.

According to the Fr\"olicher--Nijenhuis theory \cite{frn56},
condition (\ref{kr-n}) implies that the distribution of the
eigenvectors of $N$ is integrable. In application to
Darboux--Nijenhuis coordinates it means that  it is possible to
f\/ind by quadratures $2n$ functions $(x_i, y_i )$ directly from the
equa\-tions~(\ref{qv-pedr})~\cite{gz93,mag90}. Hence, the
Fr\"olicher--Nijenhuis theory allows one to construct the
Darboux--Nijenhuis variables as solutions of the equations
(\ref{qv-pedr}) or their equivalent (\ref{br-xy}).

However, for any function $f(x_i,y_i)$ equation
\[N^*df(x_i,y_i)=\lambda_i df(x_i,y_i) \]
is also satisf\/ied. Hence, any pair of Darboux--Nijenhuis
coordinates is def\/ined up to arbitrary canonical transformations.
So in the general case equations (\ref{qv-pedr}) have inf\/initely
many solutions and, thus, it is impossible to create a common
ef\/fective algorithm to solve these equations.

This problem can be partially solved. As a consequence of the
vanishing of the Nijenhuis torsion of  $N$ the eigenvalues
$\lambda_i$ always satisfy (\ref{qv-pedr})
\[
N^*d\lambda_i=\lambda_id\lambda_i .
\]
The eigenvalues $\lambda_i$ and their conjugated variables $\mu_i$
are the special Darboux--Nijenhuis coordinates \cite{fp02}, which
are distinguished because $\lambda_i$ are simply the roots of the
minimal characteristic polynomial of $N$ \bq\label{delt-N}
\Delta_N(\lambda)=\bigl(\det(N-\lambda  {\mathrm
I})\bigr)^{1/2}=\lambda^n-(c_1\lambda^{n-1}+\cdots+c_n)=\prod_{j=1}^n
(\lambda-\lambda_j) . \eq Here $c_j(\lambda_1,\ldots,\lambda_n)$
are elementary symmetric polynomials of power  $j$ which are
related with  integrals of motion $H_j$ (\ref{mag-int}) by the
Newton formulas.

The complimentary variables $\mu_j$  must be calculated as
solutions of the over\-de\-ter\-mi\-ned system of partial
dif\/ferential equations (\ref{qv-pedr}) \bq\label{eq-mu}
N^*d\mu_i=\lambda_id\mu_i,\qquad
\{\lambda_i,\mu_j\}=\delta_{ij},\qquad \{\mu_i,\mu_j\}=0 , \eq
whose solutions $\mu_j$ are still determined up to inf\/initely many
canonical  transformations $\mu_j\to \mu_j+f(\lambda_j)$.

An ef\/fective algorithm for calculating Darboux--Nijenhuis
variables has not been developed yet despite these variables play
an important role in the method of separation of variables due to
the following theorem.
\begin{theorem}[\cite{fmp01,fp02}]
If $M$ is a $2n$-dimensional $\omega N$-manifold such that in the
neigh\-bor\-hood of any point $z\in M$ recursion operator $N$ has
$n$ different functionally independent eigenvalues and if
$\{H_1,\ldots,H_n\}$ is a family of independent functions on $M$,
then the following statements are equivalent:
\begin{enumerate}\itemsep=0pt
    \item[{\rm 1)}] functions $\{H_1,\ldots,H_n\}$ are in bi-involution
    \eqref{bi-ham};
    \item[{\rm 2)}] the Lagrangian foliation $\mathcal F$ defined by $\{H_1,\ldots,H_n\}$
          is separable in Darboux--Nijenhuis coordinates;
    \item[{\rm 3)}]  the distribution $\mathcal D$ tangent to the foliation defined by $\{H_1,\ldots,H_n\}$
     is Lagrangian with respect to $\omega=P^{-1}$ and invariant with respect to $N$.
\end{enumerate}
\end{theorem}

It easy to prove that for integrable after Liouville
bi-Hamiltonian systems with the integrals of motion $H_j$
(\ref{mag-int}) recursion operator  $N$ has exactly $n$
functionally independent eigenvalues, because
$H_j=j^{-1}\sum\limits_{i=1}^n \lambda_i^j$ and
\[
dH_1\wedge\cdots\wedge dH_n=\prod_{i\neq
j}(\lambda_i-\lambda_j)d\lambda_1\wedge\cdots\wedge d\lambda_n.
\]
For bi-Hamiltonian systems integrals of motion $H_i$
(\ref{mag-int}) def\/ine a distinguished bi-Lagrangian foliation,
called principal foliation~\cite{fp02}.

For generic bi-integrable systems invariance of the distribution
$\mathcal D$ with respect to  $N$ means that there is a control
matrix $F$ with eigenvalues $(\lambda_1,\ldots,\lambda_n)$ such
that \bq\label{F-pedr} N^*dH_i=\sum_{k=1}^n F_{ij}dH_j,\qquad
i=1,\ldots,n. \eq
 In the Darboux--Nijenhuis coordinates equations
(\ref{F-pedr}) may be considered as the known Levi-Civita
criterion for separability \cite{lc04}, see \cite{fp02}.

In particular case for  bi-Hamiltonian systems (\ref{len-mag}) the
control matrix $F$ has the following form \bq\label{f-num}
F=\left(%
\begin{matrix}
  0 & 1 & 0 &\cdots & 0 \\
  0 & 0 & 1 & \cdots &0 \\
  \vdots &  & \cdots &0 & 1 \\
  c_n & c_{n+1} & \cdots & &c_1
\end{matrix}%
\right). \eq Here $c_k$ are coef\/f\/icients of the characteristic
polynomial $\Delta_N(\lambda)$ (\ref{delt-N}) of the recursion
operator~$N$.

\begin{example}
According to \cite{das89, fern93}, for a two-particle open Toda
chain recursion operator reads as
\[
N=\left(
    \begin{array}{cccc}
      p_1 & 0 & 0 & 1 \\
      0 & p_2 & -1 & 0 \\
      0 & -e^{q_1-q_2} & p_1 & 0 \\
      e^{q_1-q_2} & 0 & 0 & p_2
    \end{array}
  \right) .
\]
Integrals of motion  (\ref{tod2-int}) are reproduced by $N$
(\ref{mag-int}). Therefore eigenvalues of $N$
\[
\lambda_{1,2}=\frac{p_1+p_2}2+\frac{\sqrt{p_1^2+p_2^2-2p_1p_2+4
\exp(q_1-q_2)}}{2} .
\]
are a half of the special Darboux--Nijenhuis coordinates for the
open Toda lattice. Of course, these coordinates are the action
variables $\dot{\lambda}_i=0$.

The complimentary variables $\mu_{1,2}$  must be calculated as
solutions of the over\-de\-ter\-mi\-ned system of partial
dif\/ferential equations (\ref{eq-mu}). However  even in this simple
case we could not directly solve these thirteen PDEs (\ref{eq-mu})
for the two unknown functions $\mu_i(q_1,q_2,p_1,p_2)$.

Nevertheless we can  f\/ind variables $\mu_i$ by using the
action-angle variables (\ref{tod2-act+}), (\ref{tod2-ug}) obtained
before. Since \bq\label{tod2-act2}
\lambda_i=\frac{I_1\pm\sqrt{I_2}}{2} , \eq the second half of the
special Darboux--Nijenhuis coordinates reads as
\bq\label{tod2-ug2} \mu_i=w_1\pm 2\sqrt{I_2} w_2=\frac{q_1-q_2}2
\mp\mathrm{arctanh}\left(\frac{p_1-p_2}{\sqrt{(p_1-p_2)^2+4\exp(q_1-q_2)}}\right)
. \eq By def\/inition  action variables $\lambda_i$ are roots of the
minimal characteristic polynomial
\[
\Delta_N(\lambda)=\lambda^2-(p_1+p_2)\lambda+p_1p_2-e^{q_1-q_2},
\]
whereas angle  variables $\mu_i$ can be def\/ined in the following
way:
\[
\mu_i=\ln B(\lambda_i),\qquad B(\lambda)=-e^{q_2}(\lambda-p_1) .
\]
\end{example}
Summing up, to build the special Darboux--Nijenhuis variables we
de-facto had to use another set of separated variables $v_i$,
$u_i$ which are not Darboux--Nijenhuis variables.

In the next section we discuss how analogous variables $u$ and $v$
are related with the special Darboux--Nijenhuis variables for
$n$-particle open Toda lattice.

\section{Open Toda lattice}
Let us consider   open Toda associated with the root system of
$\mathscr A_n$ type. The Hamilton function is equal to
\begin{gather*}
H=\dfrac12\sum_{i=1}^n {p_i}^2+\sum_{i=1}^{n-1} e^{q_i-q_{i+1}}.
\end{gather*}
Here $p$, $q$ are Darboux coordinates on the manifold $M\simeq
\mathbb R^{2n}$ \bq\label{darb} \{q_i,p_j\}=\delta_{ij},\qquad
\{p_i,p_j\}=\{q_i,q_j\}=0 . \eq Bi-Hamiltonian structure of Toda
chains was investigated both in terms of  physical variables
$(p,q)$ and in terms of so-called Flaschka variables \cite{fl76}.
We will use  original physical variables $p$ and $q$, for which
second Poisson tensor has the form \cite{das89,fern93}
\bq\label{toda-gen} P^{ \prime}=\sum_{i=1}^{n-1}
e^{q_i-q_{i+1}}\dfrac{\partial}{\partial
p_{i+1}}\wedge\dfrac{\partial}{\partial p_{i}} +\sum_{i=1}^n
p_i\dfrac{\partial}{\partial q_{i}}\wedge\dfrac{\partial}{\partial
p_{i}}+\sum_{i<j}^n \dfrac{\partial}{\partial
q_{j}}\wedge\dfrac{\partial}{\partial q_{i}}. \eq Throughout the
rest of the paper $N$ is recursion operator $N=P'P^{-1}$ for the
open Toda lattice, where $P'$ is given by (\ref{toda-gen}) and $P$
is canonical tensor associated with canonical Poisson
brackets~(\ref{darb}) in $\mathbb R^{2n}$.

\subsection{The Moser method}
In this section we will give a brief review of a method for
integrating equations of motion for an open Toda lattice proposed
by Moser \cite{mos75}. A contemporary algebraic-geometric review
of the Moser method can be found in \cite{van03}.

Let us start with a $n\times n$ Lax matrix $L$ with the components
\bq\label{jac-Mat} L_{jk}=p_j \delta_{j,k}+e^\frac{q_j-q_{j+1}}{2}
(\delta_{j,k+1}+\delta_{j-1,k}) . \eq Following \cite{mos75} we
def\/ine the Weyl function \bq\label{w-fun0}
\mathcal{X}(\lambda)=\bigl( R(\lambda) \vec\alpha, \vec\alpha
\bigr),\qquad \vec\alpha=(0,0,\ldots,e^{q_n/2}), \eq where
$R(\lambda)=(L-\lambda I)^{-1}$ is a resolvent of the Jacobi
matrix (\ref{jac-Mat}) and $\vec{\alpha}$ is dynamical
normalization of the corresponding Baker--Akhiezer function. It
is known since the work of Stiltjies \cite{stil}, the Weyl
function  plays the key role in reconstruction of the matrix $L$
from its spectral data  and it is a ratio of the two monic
polynomials \bq\label{w-fun}
\mathcal{X}(\lambda)=\frac{B(\lambda)}{A(\lambda)}, \eq
 where $B(\lambda)$ is a polynomial of degree $(n-1)$ and
 $A(\lambda)=\det(L-\lambda I)$ is a polynomial of degree~$n$ with
distinct roots $\lambda_j$ \bq\label{ab-skl}
A(\lambda)=\prod_{i=1}^n(\lambda-\lambda_j),\qquad B(\lambda)=-
e^{q_n} (\lambda^{n-1}+b_2\lambda^{n-2}+\cdots+b_0) . \eq
\begin{proposition}
For the open Toda lattice two Poisson brackets associated with
tensors $P=P^{(0)}$ \eqref{mult-poi}  and $P' = P^{(1)}$ \eqref{toda-gen}
form quadratic algebras for the polynomials $A(\lambda)$,
$B(\lambda)$ and for the Weyl function $\mathcal X(\lambda)$.
\end{proposition}
It is easy to prove that at $k=0,1$ these quadratic brackets are
equal to
 \begin{gather}\label{fay-geh}
 \{A(\lambda),A(\mu)\}^{(k)}=\{B(\lambda),B(\mu)\}^{(k)}=0,\\
 \{A(\lambda),B(\mu)\}^{(k)}=\frac{\mu^kA(\lambda)B(\mu)-\lambda^kA(\mu)B(\lambda)}{\lambda-\mu}\nn
 \end{gather}
and \bq\label{ath-br}
 \left\{\mathcal{X}(\lambda),\mathcal{X}(\mu)\right\}^{(k)}=
 \frac{\bigl(\mathcal{X}(\lambda)-\mathcal{X}(\mu)\bigr)
 \bigl(\mu^k\mathcal{X}(\lambda)-\lambda^k\mathcal{X}(\mu)\bigr)}{\lambda-\mu} .
  \eq
At $k=0$ brackets (\ref{fay-geh}) are the part of the Sklyanin
brackets \cite{skl85a} and  nontrivial expression in the right
hand side of (\ref{fay-geh}) is called a Bezoutian of polynomials
$A$ and $B$ \cite{krn36}.

At $k=0$ brackets (\ref{ath-br}) give  the Atiyah--Hitchin Poisson
structure in the space of meromorphic maps
$\mathcal{X}(\lambda):\mathbb{CP}^1\to\mathbb{CP}^1$ \cite{ath88}.

\begin{remark}
According to \cite{fayg00}  the Poisson brackets between
polynomials $A(\lambda)$ and  $B(\mu)$  associa\-ted with Poisson
tensors $P^{(k)}$ (\ref{mult-poi}) at $k=0,1,\ldots,n$ are equal
to
\begin{gather}
\{A(\lambda),B(\mu)\}^{(k)}=\frac{A(\lambda)B^{[k]}(\mu)-A(\mu)B^{[k]}(\lambda)}{\lambda-\mu}
\label{fayg2}\\
\phantom{\{A(\lambda),B(\mu)\}^{(k)}}{}=  \frac{\mu^k
A(\lambda)B(\mu)-\lambda^k A(\mu)B(\lambda)}{\lambda-\mu}
+A(\lambda)A(\mu)
\frac{\bigl(\beta^{[k]}(\lambda)-\beta^{[k]}(\mu)\bigr)}{\lambda-\mu}
.\nn
\end{gather}
The right hand side of  (\ref{fayg2}) contains either remainder
\bq\label{fayB} B^{[k]}(\lambda)=\lambda^kB(\lambda)\quad\mbox{\rm
mod}\, A(\lambda), \eq either result of simple division in the
space of polynomials \bq\label{faybeta}
\beta^{[k]}=\frac{\lambda^kB(\lambda)}{A(\lambda)}, \eq which is
polynomial part of the Laurent decomposition of the ratio
$\lambda^kB(\lambda)/{A(\lambda)}$. It is easy to see that
$\beta^{[1]}=0$ and $\beta^{[1]}=-e^{q_n}$ and from (\ref{fayg2})
one gets (\ref{fay-geh}).
\end{remark}

In the Moser approach to the open Toda lattice \cite{mos75} we can
introduce the action-angle variables using the Weyl function
(\ref{w-fun0}). Namely, action coordinates are poles $\lambda_i$
(\ref{ab-skl}) of the Weyl function $\mathcal X(\lambda)$, whereas
angle variables are given by
\[
\mu_j=\ln B(\lambda_j) .
\]
It is easy to check that polynomial $A(\lambda)=\det(L-\lambda I)=
\Delta_N(\lambda)$ is a minimal characteristic polynomial of the
recursion operator $N$. Moreover, it follows from (\ref{fay-geh})
that  $\lambda_i$ and $\mu_j$ satisfy the necessary relations
(\ref{br-xy})
\[\{\lambda_i,\mu_j\}=\delta_{ij},\qquad
\{\lambda_i,\mu_j\}'=\lambda_i\delta_{ij}.
\]
We could avoid calculating of the second  Poisson brackets between
$A$ and $B$ by means of the equation \bq\label{toda-neq}
N^*d\Delta_N(\lambda)=\lambda
d\Delta_N(\lambda)+\Delta_N(\lambda)dc_1, \eq which can be
obtained from the recurrent Lenard--Magri relations
(\ref{len-mag}) rewritten in the form $N^*dH_k=dH_{k+1}$ and from
the Newton formulas connecting $H_k$ with coef\/f\/icients $c_k$
\cite{fp02}. Combining this equation with a result of \cite{ts01d}
\bq\label{toda-CC} B(\lambda)=-e^{q_n}\frac{\partial}{\partial
c_1}\Delta_N(\lambda) \eq one can easily prove that $B(\lambda)$
is a St\"ackel function \cite{fp02}
\[N^*dB(\lambda)=\lambda dB(\lambda)+\Delta_N(\lambda) de^{q_n}\]
and, thus,
\[N^*d\mu_i=\lambda_id\mu_i .\]

Summing up, the Moser variables are the special Darboux--Nijenhuis
variables for the open Toda lattice. The corresponding equations
of motion have the form
\[
\{H_i,\lambda_j\}=\partial_{\tau_i}\lambda_j=0,\qquad
\{H_i,\mu_j\}=\partial_{\tau_i}\mu_j=\lambda_j^{i-1} .
\]
Evolution of the variables $\mu_j$ with respect to the times
$\tau_i$ conjugated to the bi-Hamiltonian integrals of motion
$H_i$ (\ref{mag-int}) is linear.

Brackets(\ref{ath-br}) for the Weyl function are invariant with
respect to linear-fractional transformations \bq\label{inv-X}
\mathcal X\to\mathcal X'= \frac{a\mathcal X+b}{c \mathcal X+d} \eq
and, therefore, we can introduce another family of the separated
variables using the same Weyl function~\cite{van03}. This second
coordinate system is considered in the next section in framework
of the classical $r$-matrix theory.

\subsection{The Sklyanin method}
In this section we brief\/ly discuss an application of the generic
Sklyanin method \cite{skl85a,skl95} to the open Toda lattice. Let
us consider a $2\times 2$ monodromy matrix \bq
T(\lambda)=\left(\begin{array}{cc}
A& B \\
C& D
\end{array}\right)(\lambda)=L_1(\lambda)\cdots L_{n-1}(\lambda) L_n(\lambda), \label{22toda}
\eq where
\[
L_i=\left(\begin{array}{cc}
 \lambda-p_i &  -e^{q_i} \\
  e^{-q_i}& 0
\end{array}\right) .
\]
Monodromy matrix  $T(\lambda)$ (\ref{22toda}) is the Lax matrix
for periodic Toda lattice \cite{skl85a}, whereas  the Lax matrix
for open Toda lattice is equal to
\[
T_{o}(\lambda)=KT(\lambda)=\left(\begin{array}{cc}
A& B \\
0& 0
\end{array}\right)(\lambda),\qquad K=\left(\begin{array}{cc}
1& 0 \\
0& 0
\end{array}\right) .
\]
The entries $A$ and $B$ of this $2\times 2$ Lax matrix coincide
with
 denominator and numerator of the Weyl function (\ref{w-fun}) respectively.

\begin{proposition}
The Poisson brackets between entries of the monodromy matrix
$T(\lambda)$ \eqref{22toda} associated with Poisson tensors
$P^{(k)}$ \eqref{mult-poi} at $k=0,1$ are equal to
\begin{gather}
\big\{
\on{T(\lambda)}{1},\on{T(\mu)}{2}\big\}^{(k)}=r_{12}(\lambda,\mu)
\on{T(\lambda)}{1}\on{T(\mu)}{2}-
\on{T(\lambda)}{1}\on{T(\mu)}{2}r_{21}(\lambda,\mu)\nn\\
\phantom{\big\{\on{T(\lambda)}{1},\on{T(\mu)}{2}\big\}^{(k)}=}{}+\on{T(\lambda)}{1}s(\lambda,\mu)\on{T(\mu)}{2}-
\on{T(\lambda)}{2}s(\lambda,\mu)\on{T(\mu)}{1} , \label{br-skl2}
\end{gather}
where f\/irst $r$-matrix is equal to
\[
r_{12}(\lambda,\mu)=\dfrac{-1}{\lambda-\mu}\left(
                       \begin{array}{cccc}
                         1 & 0 & 0 & 0 \\
                         0 & 1-\frac{\lambda^k+\mu^k}{2} & \mu^k & 0 \\
                         0 & \lambda^k & 1-\frac{\lambda^k+\mu^k}{2} & 0 \\
                         0 & 0 & 0 & 1
                       \end{array}
                     \right),\qquad r_{ij}(\lambda,\mu)=\Pi r_{ji}(\lambda,\mu)\Pi,\]
whereas the second matrix $s(\lambda,\mu)$ reads as
\[s(\lambda,\mu)=\dfrac{-1}{\lambda-\mu}\left(
                       \begin{array}{cccc}
                         0 & 0 & 0 & 0 \\
                         0 & \frac{\lambda^k-\mu^k}{2} & 0 & 0 \\
                         0 & 0 & \frac{\lambda^k-\mu^k}{2} & 0 \\
                         0 & 0 & 0 & 0
                       \end{array}
                     \right).
\]
Here,  ${\on{T}{1}}(\lambda)= T(\lambda)\otimes I$,
${\on{T}{2}}(\mu)=I\otimes T(\mu)$ and $\Pi$ is a permutation
matrix in auxiliary space, i.e. $\Pi X\otimes Y=Y\otimes X\Pi$
for arbitrary matrices $X,Y$
\end{proposition}
At $k=0$ f\/irst matrix $r(\lambda,\mu)=-(\lambda-\mu)^{-1}\Pi$ is
the standard rational $r$-matrix, while the second matrix
$s(\lambda,\mu)$ is equal to zero. In this case quadratic brackets
(\ref{br-skl2}) are the Sklyanin bracket \cite{skl85a}.

\begin{proposition}
In generic case the Poisson brackets $\{\cdot,\cdot\}^{(k)}$
associated with Poisson tensors $P^{(k)}$ \eqref{mult-poi} at
$k=0,1,\ldots,n$ are equal to:
\begin{gather*}
\{A(\lambda),A(\mu)\}^{(k)}
=\{B(\lambda),B(\mu)\}^{(k)}=\{C(\lambda),C(\mu)\}^{(k)}=\{D(\lambda),D(\mu)\}^{(k)}=0 ,\\
\{A(\lambda),B(\mu)\}^{(k)}=\frac{\mu^k A(\lambda)B(\mu)-\lambda^k
A(\mu)B(\lambda)}{\lambda-\mu} +A(\lambda)A(\mu)
\frac{\bigl(\beta^{[k]}(\lambda)-\beta^{[k]}(\mu)\bigr)}{\lambda-\mu}
 ,\\
\{A(\lambda),C(\mu)\}^{(k)}=-\frac{\mu^k
A(\lambda)C(\mu)-\lambda^k A(\mu)C(\lambda)}{\lambda-\mu}
-A(\lambda)A(\mu)
\frac{\bigl(\gamma^{[k]}(\lambda)-\gamma^{[k]}(\mu)\bigr)}{\lambda-\mu}
 ,\\
\{D(\lambda),B(\mu)\}^{(k)}=-\frac{\mu^k\bigl(D(\lambda)B(\mu)-D(\mu)B(\lambda)\bigr)}{\lambda-\mu}
-\beta^{[k-1]}D(\lambda)A(\mu) ,\\
\{D(\lambda),C(\mu)\}^{(k)}=\frac{\mu^k\bigl(D(\lambda)C(\mu)-D(\mu)C(\lambda)\bigr)}{\lambda-\mu}
+\gamma^{[k-1]}D(\lambda)A(\mu) ,\\
\{B(\lambda),C(\mu)\}^{(k)}=-\frac{\mu^k
A(\lambda)D(\mu)-\lambda^k A(\mu)D(\lambda)}{\lambda-\mu}  ,
\end{gather*}
and
\begin{gather*} \{A(\lambda),D(\mu)\}^{(k)}={\frac{\lambda^k\bigl(
C(\lambda)B(\mu)- C(\mu)B(\lambda)\bigr)}{\lambda-\mu}}\\
\phantom{\{A(\lambda),D(\mu)\}^{(k)}=}{}-
\frac{A(\lambda)}{\lambda-\mu}\Bigl(B(\mu)\bigl(\beta^{[k]}(\lambda)-\beta^{[k]}(\mu)\bigr)-
C(\mu)\bigl(\gamma^{[k]}(\lambda)-\gamma^{[k]}(\mu)\bigr)\Bigr) .
\end{gather*}
Here $\beta^{[k]}=\lambda^kB(\lambda)/A(\lambda)$ \eqref{faybeta}
and  $\gamma^{[k]}=\lambda^kC(\lambda)/A(\lambda)$ are polynomial
parts of the Laurent decompositions of quotients of the
corresponding polynomials.
\end{proposition}
So, at $k>1$ we have to add to the matrices $r_{1,2}$ and $s$
dynamical terms proportional to the functions $\beta^{[k]}$  and
$\gamma^{[k]}$.

According to \cite{fl76,skl85a, skl95} the  separated coordinates
are poles of the corresponding Baker--Akhiezer function with the
standard normalization $\vec\alpha=(0,1)$. In this case the f\/irst
half of variables are coming from $(n-1)$ f\/inite roots and
logarithm of leading coef\/f\/icient of the non-diagonal entry of the
monodromy matrix \bq\label{B-int}
B(\lambda)=-e^{-v_n}\prod_{j=1}^{n-1}(\lambda-u_j) ,\qquad
v_n=-q_n , \eq Another half is given by
 \bq \label{toda-per} v_j=-\ln
A(u_j),\quad j=1,\ldots,n-1,\qquad\mbox{\rm and}\qquad
u_n=-c_1=\sum_{i=1}^n p_i . \eq In the separated variables
polynomial $A(\lambda)$ reads as \bq\label{A-int}
A(\lambda)=\left(\lambda+\sum_{j=1}^{n}u_j\right)\prod_{j=1}^{n-1}(\lambda-u_j)+
\sum_{j=1}^{n-1} e^{-v_j}\prod_{i\neq
j}^{n-1}\dfrac{\lambda-u_i}{u_j-u_i} . \eq This def\/inition of the
separated variables is obviously related with the following
transformation $\mathcal X\to\mathcal X^{-1}$ of the Weyl function
\cite{van03}.

It follows from (\ref{br-skl2}) that at $k=0,1$ \bq\label{br-AC}
\{A(\lambda),B(\mu)\}^{(k)}=\dfrac{1}{\lambda-\mu}\Bigl(\mu^kA(\lambda)B(\mu)-\lambda^kA(\mu)B(\lambda)\Bigr)
. \eq Substituting these brackets into the equations
\[
0=\{A(\lambda),B(u_j)\}^{(k)}=\left.\{A(\lambda),B(\mu)\}^{(k)}\right|_{\mu=u_j}
+B'(u_j) \{A(\lambda),u_j\}^{(k)}
\]
one gets \bq\label{toda-eqm} \{A(\lambda),u_j\}^{(k)}=\lambda^k
A(u_j) \prod_{i\neq j}^{n-1}\dfrac{\lambda-u_i}{u_j-u_i} ,\qquad
j=1,\ldots,n-1. \eq If $\lambda=u_i$  this implies  part of the
necessary relation (\ref{br-xy})
\[\{u_j,v_i\}^{(k)}=u_i^k \delta_{ij},\qquad k=0,1 ,\qquad i,j=1,\ldots,n-1.\]
Now we can collect all the coef\/f\/icients with the same  powers of
$\lambda$ and $\mu$ in (\ref{br-AC}) and prove that
\begin{alignat*}{4}
&\{u_n,v_n\}=1,\qquad && \{u_i,u_n\}=0,\qquad&& \{v_n,v_i\}=0 ,&\\
&\{u_n,v_n\}'=- \sum_{i=1}^{n} u_i, \qquad&& \{u_n,u_i\}'=f_i=
\frac{e^{-v_i}}{\prod\limits_{j\neq i}^{n-1}(u_i-u_j)},\qquad&&
\{v_n,v_i\}'=-1 .&
\end{alignat*}
In a similar manner from $\{B(\lambda),B(\mu)\}^{(k)}=0$ one gets
\[\{u_i,u_j\}^{(k)}=\{v_n,u_j\}^{(k)}=0 ,\qquad i,j=1,\ldots,n-1,\]
and from $\{A(\lambda),A(\mu)\}^{(k)}=0$ it follows that
\begin{alignat*}{3}
&\{u_n,v_i\}=0,\qquad &&\{v_i,v_j\}=0 ,\qquad  i,j=1,\ldots,n-1,&\\
\\
&
\{u_n,v_i\}'=g_i=\left.-e^{-v_n}\frac{A'(\lambda)}{B'(\lambda)}\right|_{\lambda=u_i},
\qquad &&\{v_i,v_j\}'=0 .&
\end{alignat*}
Here $A'(\lambda)$ and $B'(\lambda)$ are derivatives by $\lambda$.

Summing up, the $2n$ separated variables $v_i$ and $u_i$ are the
Darboux variables
\[\omega=\sum_{i=1}^n du_i\wedge dv_i,\]
but the corresponding recursion operator $N$ consists of two
diagonal and three non-diagonal terms
\begin{gather}
 N=\sum_{i=1}^{n-1}u_i\left(
\dfrac{\partial }{\partial  u_i}\otimes du_i+ \dfrac{\partial
}{\partial v_i}\otimes dv_i\right)-\sum_{i=1}^{n}u_i\left(
\dfrac{\partial }{\partial  u_n}\otimes dv_n+ \dfrac{\partial
}{\partial v_n}\otimes du_n\right)\nn\\
\phantom{N=}{}+\sum_{i=1}^{n-1}f_i\left(\dfrac{\partial }{\partial
u_i}\otimes du_n-\dfrac{\partial }{\partial u_n}\otimes
du_i\right) +\sum_{i=1}^{n-1}\left( \dfrac{\partial }{\partial
v_n}\otimes
dv_i- \dfrac{\partial }{\partial v_i}\otimes dv_n\right)\nn\\
\phantom{N=}{}+ \sum_{i=1}^{n-1}g_i\left(\dfrac{\partial
}{\partial v_i}\otimes du_n+\dfrac{\partial }{\partial u_n}\otimes
dv_i\right) .\label{N-uv}
\end{gather}
Thus $2n$ separated variables $v_i$ and $u_i$ obtained in
framework of the Sklyanin method \cite{skl85a} are not the
Darboux--Nijenhuis variables. Evidently it is related with the
dif\/ference in the form of the corresponding separated equations,
which follow directly from the def\/initions of $v_j$
and~$u_n$~(\ref{toda-per})
\begin{gather}\label{tod-seq}
\Phi_j=e^{-v_j}-\Delta(u_j,\alpha_1,\ldots,\alpha_n)=0 ,\qquad
\Phi_n=u_n+\alpha_1=0 .
\end{gather}
The f\/irst $(n-1)$ equations of motion, see (\ref{toda-eqm}), are
linearized by the Abel transformation~\cite{skl85a,smirn98}
\[\left\{A(\lambda),\sum_{k=1}^{n-1} \int^{v_k} \sigma_{j}\right\}=-\lambda^{j-1},\qquad
\sigma_j=\frac{\lambda^{j-1} d\lambda }{\Delta_N(\lambda)},\qquad
j=1,\ldots,n-1 ,
\]
where $\{\sigma_j\}$ is a basis of Abelian dif\/ferentials of f\/irst
order on an algebraic curve $z=\Delta_N(\lambda)$ corresponding to
separated equations (\ref{tod-seq}).

It means that we can introduce the action-angle variables
\[
I_j=c_j,\qquad w_1=v_n,\qquad\mbox{\rm and} \qquad
w_{j+1}=\sum_{k=1}^{n-1}\int^{v_k} \sigma_{j} ,
\]
such that evolution of variables $w_j$ with respect to times
$\tau_j$ conjugated to the action
 variables~$I_j$ are linear. These
action variables are related with the previous ones
\[
A(\lambda)=\prod_{i=1}^n(\lambda-\lambda_i)=\lambda^n-\sum_{j=1}^{n}
I_j\lambda^{n-j} .
\]
Bi-Hamiltonian integrals of motion $H_j$ (\ref{mag-int}) produce
the f\/lows which preserve the spectrum of the $n\times n$ Jacobi
matrix $L$ (\ref{jac-Mat}). Integrals of motion $I_j=c_j$ produce
the transversal  to the isospectral manifolds f\/lows, which
preserve the divisor~\cite{van03}.

As sequence the special Darboux--Nijenhuis variables $\lambda$,
$\mu$ are dual to the Sklyanin variab\-les~$u$,~$v$. Namely,
$\lambda_i$, $\mu_i$ are roots of polynomial $A(\lambda)$ and
values of polynomial $B(\lambda)$ at $\lambda=\lambda_i$, while
$u_j$, $v_j$ are roots of polynomial $B(\lambda)$ and values of
polynomial $A(\lambda)$ at $\lambda=u_j$.

\begin{remark}
From the factorization of the monodromy matrix $T(\lambda)$
(\ref{22toda}) one gets
\[
 B_{n}(\lambda)=-e^{q_{n}}A_{n-1}(\lambda) \qquad \Rightarrow\qquad B_{n}(u_j)=-e^{q_{n}}A_{n-1}(\lambda_j)=0.
\]
This implies that the Moser variables $\lambda_j$ for a
$(n-1)$-particle chain, i.e. special Darboux--Nijenhuis variables,
coincide with the Sklyanin variables $u_j$, $i=1,\ldots,n-1$ for a
$n$-particle chain.

We can prove this fact directly using matrix representation of the
recursion operator for $(n-1)$-particle chain, which  can be
obtained by the matrix rep\-re\-sen\-tation (\ref{N-uv}) deleting
the $n$-th and $2n$-th rows and columns.
\end{remark}

\begin{remark}
The Sklyanin variables are the separated variables for open and
periodic Toda lattices simultaneously, in contrast with the Moser
variables. For the periodic Toda lattice the Darboux--Nijenhuis
variables were constructed by using Flaschka variables
in~\cite{fmp01}.
\end{remark}

\begin{example}
At $n=2$ the Sklyanin variables are equal to
\[
v_1 = -\ln(-e^{q_1-q_2}),\qquad v_2 = -q_2, \qquad u_1 =
p_1,\qquad u_2 = p_1+p_2 ,
\]
and
\[ N=\left(\begin{array}{cccc} u_1& 0& 0& -e^{-v_1}\\
2u_1-u_2\quad& u_2-u_1\quad& e^{-v_1}\quad& 0\\
0& -1& u_1& 2u_1-u_2\\
1& 0& 0& u_2-u_1
\end{array}\right).
\]
We have three dif\/ferent families of the separated variables for
the Toda lattice at $n=2$ only.
\end{example}

\section{Conclusion}
For the open Toda lattice associated with the root system of
$\mathscr A_n$ type we prove that the Moser variables are special
Darboux--Nijenhuis variables, while the Sklya\-nin variables are
``almost Darboux--Nijenhuis variables'', in which recursion
operator consists of a diagonal part and two non-diagonal rows and
columns only.

The similar results for the generalized open Toda lattices
associated with the root systems of $\mathscr B_n$, $\mathscr C_n$
and $\mathscr D_n$ type were found in~\cite{tsnew}.

Nevertheless we have to underline that construction of the Moser
variables and the Sklyanin variables directly in the framework of
the bi-Hamiltonian geometry is an open question.

\subsection*{Acknowledgements}
We would like to thank I.V.~Komarov and V.I.~Inozemtsev  for
useful and interesting discussions. The research was partially
supported by the RFBR grant 06-01-00140.

\LastPageEnding
\end{document}